\documentclass{emulateapj}
\usepackage{apjfonts}
\def\arcsec{$\,^{\prime\prime}$~}

\newcommand{\be}{\begin{equation}}
\newcommand{\bel}[1]{\begin{equation}\label{eq:#1}}
\newcommand{\ee}{\end{equation}}
\newcommand{\bd}{\begin{displaymath}} 
\newcommand{\ed}{\end{displaymath}}   
\newcommand{\bea}{\begin{eqnarray}}
\newcommand{\beal}[1]{\begin{eqnarray}\label{eq:#1}}
\newcommand{\eea}{\end{eqnarray}}

\newcommand{\eqref}[1]{\ref{eq:#1}}

\newcommand{\lsim }{{\lower0.8ex\hbox{$\buildrel <\over\sim$}}}
\newcommand{\gsim }{{\lower0.8ex\hbox{$\buildrel >\over\sim$}}}

\def\simge{\mathrel{%
   \rlap{\raise 0.511ex \hbox{$>$}}{\lower 0.511ex \hbox{$\sim$}}}}
\def\simle{\mathrel{
   \rlap{\raise 0.511ex \hbox{$<$}}{\lower 0.511ex \hbox{$\sim$}}}}

\newcommand{\Msun}{\ifmmode {M_{\odot}}\else${M_{\odot}}$\fi}
\newcommand{\Lsun}{\ifmmode {L_{\odot}}\else${L_{\odot}}$\fi}
\newcommand{\Rsun}{\ifmmode {R_{\odot}}\else${R_{\odot}}$\fi}

\shorttitle{Thermal X-ray Radiation from SAX J1808.4-3658}
\shortauthors{Heinke et al.}

\begin{document}
\title{Constraints on Thermal X-ray Radiation from SAX J1808.4--3658 and Implications for Neutron Star Neutrino Emission\altaffilmark{1}}  

\author{C. O. Heinke\altaffilmark{2,3}, P.~G. Jonker\altaffilmark{4,5,6}, R. Wijnands\altaffilmark{7}, and R.~E. Taam\altaffilmark{2}}

\altaffiltext{1}{Based on observations obtained with XMM-Newton, an ESA science mission with instruments and contributions directly funded by ESA Member States and NASA}

\altaffiltext{2}{Northwestern University, Dept. of Physics \&
  Astronomy, 2145 Sheridan Rd., Evanston, IL 60208;
cheinke@northwestern.edu}

\altaffiltext{3}{Lindheimer Postdoctoral Fellow}
\altaffiltext{4}{SRON, Netherlands Institute for Space Research, Sorbonnelaan 2, 3584~CA, Utrecht, the Netherlands}
\altaffiltext{5}{Harvard--Smithsonian  Center for Astrophysics, 60 Garden Street, Cambridge, MA~02138, Massachusetts,
U.S.A.}
\altaffiltext{6}{Astronomical Institute, Utrecht University, PO Box 80000, 3508 TA, Utrecht, the Netherlands}

\altaffiltext{7}{Astronomical Institute "Anton Pannekoek", University of Amsterdam, Kruislaan 403, 1098 SJ, The Netherlands}


\begin{abstract}

Thermal X-ray radiation from neutron star soft X-ray transients in 
quiescence provides the strongest constraints on the cooling rates 
of neutron stars, and thus on the interior composition and properties of 
matter in the cores of neutron stars.  We analyze new (2006) and archival 
(2001) XMM-Newton observations of the accreting millisecond pulsar 
SAX J1808.4--3658 in quiescence, which provide the most stringent constraints 
to date. The X-ray spectrum of SAX J1808.4--3658 in the 2006 observation 
is consistent with a power-law of photon index $1.83\pm0.17$, without 
requiring the presence of a blackbody-like component from a neutron star 
atmosphere. Our 2006 observation shows a slightly lower 0.5--10 keV X-ray 
luminosity, at a level of 68$^{+15}_{-13}$\% that inferred from 
the 2001 observation. Simultaneous fitting of all available XMM data allows 
a constraint on the quiescent neutron star 
(0.01--10 keV) luminosity of $L_{NS}<1.1\times10^{31}$ erg s$^{-1}$. 
This limit excludes some current models of neutrino 
emission mediated by pion condensates, and provides further evidence 
for additional cooling processes, such as neutrino emission via direct Urca 
processes involving nucleons and/or hyperons, in the cores of 
massive neutron stars.  
\end{abstract}

\keywords{binaries : X-rays --- dense matter --- neutrinos --- stars: neutron}


\section{Introduction}\label{s:intro}

The X-ray transient SAX J1808.4--3658 (hereafter 1808) has provided
many fundamental breakthroughs in the study of accreting neutron
stars (NSs). It was discovered in 1996 by {\it BeppoSAX}'s Wide Field
Cameras, and type I X-ray bursts were seen, identifying it as an
accreting NS and constraining the distance 
\citep{intZand98,Galloway06}.  Coherent millisecond X-ray
pulsations, the first discovered in accreting systems,
 were identified during an outburst using the Rossi X-ray Timing 
Explorer \citep[RXTE;][]{Wijnands98}.  
Burst oscillations have also been seen at 1808's 
401 Hz spin frequency, confirming that thermonuclear burst
oscillations in low-mass X-ray binaries (LMXBs) represent the spin
period of the NS \citep{intZand01b,Chakrabarty03}.
A pair of kilohertz quasiperiodic oscillations (QPOs) were seen from 1808,
with a frequency difference equal to one-half of the spin period,
forcing a revision of the most popular models for QPOs
\citep{Wijnands03c}.  Optical observations of the brown dwarf companion
while 1808 was in quiescence showed a sinusoidal optical modulation
attributed to heating of the companion \citep{Homer01}.  However, the
required irradiating luminosity is larger than the available X-ray
luminosity, giving rise to speculation that a radio pulsar mechanism
is active during quiescence \citep{Burderi03,Campana04b}.  

In addition to these discoveries, 1808 has provided one of the lowest
quiescent thermal luminosities yet measured from any accreting NS
\citep{Campana02,Wijnands02c}, along with 1H1905+000 \citep{Jonker06}.  
 Transiently accreting NSs in quiescence are usually seen to have soft,
blackbody-like X-ray spectra, often accompanied by
a harder X-ray component generally fit by a power-law of photon index
1--2 \citep{Campana98a}.  The harder component is of unknown origin;  
an effect of continued accretion, or a shock from a pulsar wind have 
been suggested \citep{Campana98a}.   
 The blackbody-like component is generally understood as the 
radiation of heat from the NS surface. This heat is produced by deep crustal 
heating during accretion, and is radiated by the crust 
on a timescale of $10^4$ years, producing a steady
quiescent thermal NS luminosity \citep{Brown98, Campana98a, Haensel90}. 
 Measurement of the blackbody-like component is
particularly important because it indirectly constrains the interior
structure of NSs. 

What fraction of this heat escapes the NS as neutrinos rather than photons 
depends on the physical conditions (i.e.  composition, density, and 
pressure) of the NS interior.  If the outburst 
history (fluence, recurrence time) and distance of a NS are reasonably 
well-known, then the determination of the quiescent thermal NS luminosity 
constrains the neutrino vs. photon emission, and thus models for the NS 
interior \citep{Yakovlev04,Levenfish06}. For example, the 
transient Cen X-4 has been identified as having a rather low quiescent X-ray 
luminosity compared to deep crustal heating predictions. 
This suggests that Cen X-4 has 
enhanced neutrino emission, produced in the high density core of a relatively 
high mass NS \citep{Colpi01}.  Many other LMXBs have also shown low 
quiescent thermal X-ray luminosities, indicating either 
enhanced neutrino emission or 
extremely long quiescent intervals \citep[e.g.][]{Wijnands01,Jonker04,
Tomsick04,Jonker06}.  The coolest of these provide the strongest constraints 
to date on neutrino cooling from NS cores, as a broader range of cooling 
rates is necessary to explain the data than for young cooling pulsars 
\citep{Page04,Yakovlev04}. 

 1808 is a particularly interesting system 
 due to its known distance and constrained mass transfer rate 
\citep{Bildsten01}.  \citet{Campana02} observed 
1808 in quiescence with XMM in 2001, finding  
a low luminosity ($L_X$(0.5-10 keV)$=5\times10^{31}$ ergs s$^{-1}$, for $d=2.5$ kpc) and a relatively hard spectrum, with less than 10\% of the X-ray flux  
attributable to a possible blackbody-like component.  
We have obtained a deeper XMM observation of 1808 
in quiescence in 2006 to place more stringent constraints on  
neutron star cooling processes. 

\section{Data Reduction}\label{s:obs}

We observed 1808 on September 14, 2006 (Obs-Id 0400230401) 
for 54 ksec with XMM's EPIC camera, using two MOS CCD detectors 
\citep{Turner01} with medium filters and one pn CCD detector 
\citep{Struder01} with a thin filter.   
We also downloaded the March 24, 2001 XMM observation \citep[Obs-Id 0064940101, reported by ][]{Campana02} from the HEASARC archive.
All data were reduced using FTOOLS and SAS version { 7.0.0}.  
We used only the
MOS data from 2001, since the pn data were taken in timing mode, and the 
target was too faint to be detected in this mode.  
Intervals of flaring background were excluded by excluding times when the 
total MOS count rate exceeded 5 0.2--12 keV counts per second, and times when 
the total pn count rate exceeded 50 0.2--12 keV counts per second. 
This left 36.6, 47.0, 48.1, and 39.3 ksec in the 2001 MOS data, the 2006 
MOS1 data, the 2006 MOS2 data, and the 2006 pn data, respectively.
Event grades higher than 12 were also excluded.  We
extracted spectra from a 10\arcsec circle around the position of 1808, 
correcting the fluxes for the fraction of photons collected within 
the extraction radius, and combining the pairs of simultaneous MOS 
spectra and responses using FTOOLS.  We generated 
response and effective area files using the SAS tasks {\it rmfgen} and
{\it arfgen}, and produced background spectra from  90\arcsec
circular source-free regions on the same CCD. 
The spectra were grouped to $>$15 counts per bin for the MOS data, 
and $>$30 counts per bin for the pn data (other choices gave similar results). 
To assess variability within the 2006 and 2001 observations, 
background-subtracted lightcurves were produced within 
SAS and analyzed using HEASARC's  
XRONOS software. 
KS and $\chi^2$ tests on the last 15 ksec of 0.2--12 keV pn data 
(unaffected by background flaring) revealed no evidence of variability.  

\section{Spectral Analysis}\label{s:spec}

Our fitting includes photoelectric absorption (XSPEC model 
{\it phabs}), with a hydrogen column density, $N_H$, fixed 
at the value derived from observations in outburst ($1.3\times10^{21}$ 
cm$^{-2}$; note this is equal to the value derived
from \citealt{Dickey90}).  We also tested models with photoelectric 
absorption as a free parameter, in all cases finding $N_H$ consistent with 
the outburst value.  Quoted errors are at 90\% confidence. 

We fit the combined 2001 MOS spectrum to a power-law model, 
finding $\Gamma=1.72\pm0.28$ (see Table 1).  Fits using only 
a hydrogen-atmosphere model, the NSATMOS model of \citet{Heinke06a}
(similar to
the NSA model of \citealt{Zavlin96}) gave poor fits ($\chi^2_{\nu}>4.8$).  
We then performed fits with NSATMOS plus a power-law, 
fixing the true NS radius to 10 km, the gravitational mass to 1.4 \Msun, and the distance to 3.5 kpc \citep{Galloway06}. 
No thermal component is required, but a thermal component with $kT<$42 eV 
(90\% confidence\footnote{$kT$ is the redshifted temperature, or $kT_{\infty}$}) is permitted, thus placing a limit on the NS's 
thermal 
0.01--10 keV (essentially bolometric) luminosity of $L_{NS}<2.4\times10^{31}$ erg s$^{-1}$.  
  The inclination of 
this system is known to be low \citep{Bildsten01}, and it is rare for 
LMXBs to show higher $N_H$ in quiescence than outburst \citep[e.g.][]{Jonker04c}, so we regard a 
higher $N_H$ as very unlikely.  
The total 0.5--10 keV 
unabsorbed luminosity is $L_X=7.6^{+1.7}_{-1.5}\times10^{31}$ erg s$^{-1}$.  

\begin{figure}
\figurenum{1}
\includegraphics[angle=270,scale=.36]{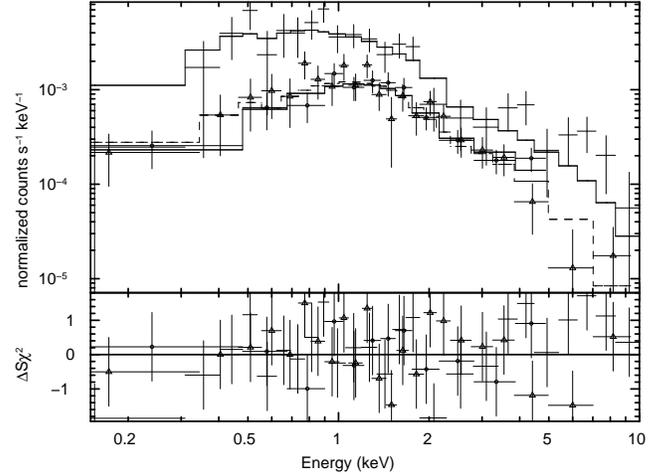}
\caption[f1.eps]{ \label{fig:spectrum}
 Top panel: XMM X-ray spectra (data and best-fit power-law model) of SAX J1808.4--3658.  
Solid line with crosses: 2006 pn data, model.  
Dashed line with triangles: 2006 MOS data, model. 
Solid line with circles: 2001 MOS data, model.  
Bottom panel: residuals to the fit in units of $\chi^2$.
} 
\end{figure}

For the 2006 data, we find a similar spectral shape, and therefore 
fit similar models to the pn and MOS data simultaneously.  
A simple power-law fits the data adequately, with a photon index of 
$\Gamma=1.83\pm0.17$ and an unabsorbed 
$L_X$(0.5--10 keV)$=5.2\pm0.7\times10^{31}$ erg s$^{-1}$.  The  
2006 flux appears less than the 2001 flux.  We test this by 
fitting the spectra simultaneously and  
tying their power-law slopes together, finding that the 
2001 0.5--10 keV unabsorbed flux is higher at 97\% confidence; 
the 2001 flux is 1.28$^{+0.24}_{-0.21}$ (90\% conf.) that of 
the 2006 observation.  If the power-law slopes are allowed to vary ($1.61\pm0.3$ for 2001, $1.83\pm0.17$ for 2006), then 
the best-fit flux ratio is 1.47$^{+0.35}_{-0.27}$.  

Fitting the 2006 and 2001 data allows a tighter constraint on the presence 
of a NS atmosphere component than the 2001 data alone, requiring a NS $kT<34$ eV and a thermal 0.01--10 keV NS 
luminosity $L_{NS}<1.1\times10^{31}$ erg s$^{-1}$. 
Choosing a NS radius of 12 km, or a mass of 2.0 \Msun, varies this 
constraint by only 3\%.  
The rather tight distance limits of \citet{Galloway06} ($3.5\pm0.1$ kpc) 
produce only a 6\% uncertainty.  
Allowing the $N_H$ to float freely permits a thermal 0.01--10 keV 
NS luminosity $L_{NS}<1.0\times10^{32}$ ergs s$^{-1}$ 
(for $N_H=1.7\times10^{21}$ cm$^{-2}$). 



\section{Ramifications}

We have estimated the time-averaged mass transfer rates for 1808, 
and several other transient LMXBs (Aql X-1, Cen X-4, 4U 1608--52, KS1731--260, RX 1709--2639, MXB 1659--29, XTE 2123--058, SAX 1810.8--2609, and those in Terzan 5 and NGC 6440), from the RXTE All-Sky Monitor (ASM) record (1996 to Nov. 2006)
 under the assumption that the time-averaged mass accretion rate 
over the last 10 years reflects the time-averaged mass transfer rate 
(Table 2).   We use PIMMS 
and a power-law of photon index 2 to convert the ASM countrates during 
outbursts into 0.1--20 keV fluxes\footnote{We have verified that this conversion is correct to within 50\% for outbursts of the transients EXO 1745-245 and Aquila X-1.}.  This is, of course, a rough approximation, 
as the spectral shapes of LMXBs in outburst vary substantially.  
Additional sources of potential error include poor ASM time coverage of some 
outbursts, uncertainty in the NS mass and radius (affecting the energy 
released per accreted gram, and thus the 
conversion from $L_X$ to mass accretion rate), variability in the 
mass transfer rate, and uncertain distances (which will equally affect 
the quiescent luminosity). We plot an arbitrary uncertainty of 50\% in both 
mass transfer rate and quiescent luminosity for each point in Fig. 2.  
For Cen X-4 we use the lowest measured quiescent luminosity, and the 
mass transfer rate limit inferred if Cen X-4 undergoes outbursts 
every 40 years with a fluence similar to its 1969 outburst \citep{Chen97}.  
The NS component flux for Aquila X-1 is somewhat uncertain and possibly variable \citep{Rutledge02b,Campana03}.  
We assume that all outbursts from NGC 6440 since 1971 have been detected. 
 For KS 1731-260 we assume 
that the average flux seen with RXTE-ASM during outburst was the average flux during the entire 12.5 year outburst. For KS 1731--260 and the transient in Terzan 1 (for which we take a 12-year outburst) we take a minimum recurrence time of 30 years. 

\begin{figure}
\figurenum{2}
\includegraphics[angle=0,scale=.45]{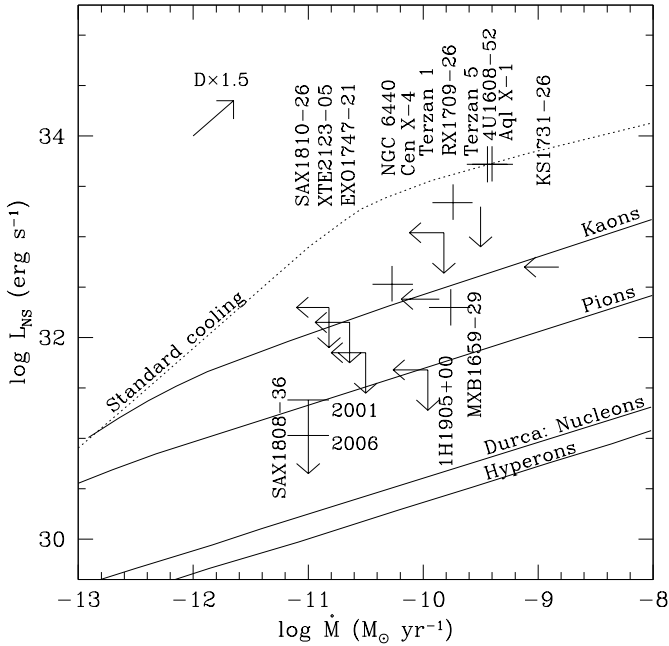}
\caption[f2.eps]{ \label{fig:cooling}
Cooling curves for various NS interior neutrino emission scenarios, compared 
with measurements (or 95\% conf. upper limits) of the quiescent 0.01--10 keV 
NS luminosity and time-averaged 
mass transfer rate for several NS transients (see Table 2).    
Cooling curves 
taken from \citet{Yakovlev04}; the dotted curve represents a low-mass NS, 
while the lower curves represent high-mass NSs with kaon or pion condensates, 
or direct Urca (Durca) processes mediated by nucleons or hyperons.   
Limits on the quiescent NS luminosity of SAX J1808.4--3658 are given for the 
2001 and 2006 observations.  The effect of 
a distance error as large as a factor 1.5 is also indicated (upper left). 
} 
\end{figure}

For 1808 we derive a time-averaged mass transfer rate of 
$1.0\times10^{-11}$ \Msun/year, an excellent match to the prediction of 
general relativity of $0.95\times10^{-11}$($M_2$/0.05 \Msun) \Msun/year 
\citep{Bildsten01}.  We note that the true mass
transfer rate cannot be less than $7\times10^{-12}$ \Msun/year for a NS mass
$\geq$1.4 \Msun\ under the assumption of an index $n=-1/3$
for the donor's mass-radius relation, or less than $3.5\times10^{-12}$
\Msun/year for an index $n=1$ mass-radius relation.  
It is unclear whether the entropy of the donor can be maintained by the 
low quiescent luminosity of the neutron star \citep{Homer01,Burderi03}; 
we plan future observations and modeling to address this issue.

We have plotted the cooling curves calculated by \citet{Yakovlev04} for 
a variety of models in Fig. 2. 
Low-mass NSs will cool slowly (dotted line in Fig.~2); in the model of \citet{Yakovlev04}, only through 
photon emission and neutron-neutron neutrino bremsstrahlung processes, 
while modified Urca neutrino emission is suppressed by proton superfluidity.   
Other slow cooling models (invoking, e.g., neutrino emission through Cooper pair formation) give similar results \citep[e.g.][]{Page04}.  
Higher-mass NSs should have 
higher central densities, sufficient to promote more rapid direct Urca 
neutrino cooling processes involving nucleons and/or hyperons, or 
direct Urca-like processes mediated by pions, kaons, or quark matter,
in their cores. Medium-mass models can produce intermediate cooling 
rates if proton superconductivity is important at low densities,
as its decay at moderate densities can allow a smooth 
transition between fast and slow cooling rates \citep{Yakovlev03,Levenfish06}.
Thus NSs of different masses 
should lie between the top curve and one of the lower curves in Fig. 2, 
where the lower curve is the maximum neutrino cooling curve.  
We note a possible trend that NSs with low mass transfer rates seem to have 
particularly low quiescent luminosities, well below the 
``standard cooling'' predictions.  This might be explained through 
binary evolution; NSs with low mass transfer rates may be very old systems,
which may have accreted significant mass.  In enhanced neutrino emission 
scenarios, these massive NSs would then have higher neutrino and lower 
photon luminosities than younger systems.  

Our constraint on the quiescent thermal 0.01--10 keV NS luminosity of 
1808 from the 2006 observations thus seems to rule out some models 
of direct Urca neutrino emission via pion condensates, favoring direct 
Urca processes involving nucleons and/or hyperons. 
An extremely large distance uncertainty of 50\% \citep[see Fig.~2; cf. ][]{Galloway06}
 would be required to bring 1808's thermal luminosity up to the pion 
condensate predictions.  
The 2001 observation, by itself, rules out some models of direct Urca neutrino 
emission from kaon condensates.  
Other modelers of NS cooling have suggested that medium effects \citep{Blaschke04} or diquark condensates \citep{Grigorian05} could provide a wide range of NS cooling rates.  These models may also be sufficient to explain the data on 1808 presented here.  
Our results agree with the principal conclusions of, e.g., \citet{Yakovlev04}, \citet{Levenfish06}, and \citet{Page06}, and provide a firmer observational basis for future studies.

\acknowledgements

We thank C.~J. Deloye, E.~F. Brown, and A.~W. Steiner for useful discussions, and the referee for a rapid and constructive report.  
RXTE ASM results provided by the ASM/RXTE teams at MIT 
and NASA's GSFC.
 COH acknowledges support from the Lindheimer Postdoctoral Fellowship at 
Northwestern University, and NASA 
XMM grant NNX06AH62G.  PGJ acknowledges support from the Netherlands 
Organization for Scientific Research. 


\bibliography{src_ref_list}
\bibliographystyle{apj}

\clearpage

\begin{deluxetable}{ccccccc}
\tablewidth{6.5truein}
\tablecaption{\textbf{Spectral Fits to SAX J1808.4--3658}}
\tablehead{
\colhead{\textbf{Epoch}}  & $N_H$ & $\Gamma$ & 
 $\chi^2_{\nu}$/dof & $L_X$  & $kT$ 
& $L_{NS}$  \\
   & ($10^{22}$ cm$^{-2}$) &  &  & (erg s$^{-1}$) & (eV) & (erg s$^{-1}$) \\
}
\startdata
2001 & (0.13) &  $1.61\pm0.3$ & 0.51/9  & 7.6$^{+1.7}_{-1.5}\times10^{31}$ & $<42$ & $<2.4\times10^{31}$  \\  
2006 & (0.13) &  $1.83\pm0.17$ & 0.86/45 & 5.2$\pm0.7\times10^{31}$ &   $<35$ & $<1.2\times10^{31}$     \\  
2001 \& 2006  & (0.13) & $1.83\pm0.16$ & 0.79/55 & 5.2$\pm0.7\times10^{31}$ &  $<34$ & $<1.1\times10^{31}$  \\  
2001 \& 2006 & 0.15$\pm0.04$ & $1.93^{+0.37}_{-0.29}$ & 0.78/54 & 5.2$\pm1.0\times10^{31}$ & $<61$ & $<1.0\times10^{32}$ \\ 
\enddata
\tablecomments{Spectral fits with power-law plus NSATMOS model to SAX J1808.4--3658.  
Errors are 90\% confidence for a single parameter.  
$N_H$ is held fixed in the first three rows. 
$L_X$ for 0.5--10 keV range, $L_{NS}$ for 0.01--10 keV. 
}
\end{deluxetable}

\begin{deluxetable}{ccccccccc}
\tablewidth{7.5truein}
\tablecaption{\textbf{Luminosities and Mass Transfer Rates}}
\tablehead{
\colhead{\textbf{Source}}  & $N_H$ & $kT$ & D & Outbursts & Years & \.{M}  &  $L_{NS}$ & Refs \\
   & ($10^{22}$ cm$^{-2}$) &  (eV) & (kpc) &  &  & (\Msun\ yr$^{-1}$) & (erg s$^{-1}$) &  \\
}
\startdata
Aql X-1 & $4.2\times10^{21}$ & $\sim$94 &  5 & 8 & 10.7 & $4\times10^{-10}$ & $5.3\times10^{33}$ & 1,2,3,4 \\
Cen X-4 & $5.5\times10^{20}$ & 76 & 1.2 & - & - & $<3.3\times10^{-11}$ & $4.8\times10^{32}$ & 5,3 \\
4U 1608--522 & $8\times10^{21}$ & 170 & 3.6 & 4 & 10.7 & $3.6\times10^{-10}$ & $5.3\times10^{33}$ & 6,3,4 \\    
KS 1731--260 & $1.3\times10^{22}$ & 70 & 7 & 1 & 30 & $<1.5\times10^{-9}$ & $5\times10^{32}$ & 7,4 \\
MXB 1659--29 & $2.0\times10^{21}$ & 55 & $\sim$10? & 2 & 10.7 & $1.7\times10^{-10}$ & $2.0\times10^{32}$ & 7,4 \\
EXO 1747--214 & $4\times10^{21}$ & $<63$ & $<11$ & - & - & $<3\times10^{-11}$ & $<7\times10^{31}$ & 8 \\
Terzan 5 & $1.2\times10^{22}$ & $<131$ & 8.7 & 2 & 10.7 & $3\times10^{-10}$ & $<2.1\times10^{33}$ & 9,10,4 \\
NGC 6440 & $7\times10^{21}$ & 87 & 8.5 & 3 & 35 & $1.8\times10^{-10}$ & $3.4\times10^{32}$  & 11,4 \\
Terzan 1 & $1.4\times10^{22}$ & 74 & 5.2 & - & - & $<1.5\times10^{-10}$ & $<1.1\times10^{33}$ & 12 \\
XTE 2123--058 & $6\times10^{20}$ & $<66$ & 8.5 & 1 & 10.7 & $<2.3\times10^{-11}$ & $<1.4\times10^{32}$ & 3,4 \\
SAX J1810.8--2609 & $3.3\times10^{21}$ & $<72$ & 4.9 & 1 & 10.7 & $<1.5\times10^{-11}$ & $<2.0\times10^{32}$ & 13,3,4 \\
RX J1709--2639 & $4.4\times10^{21}$ & 122 & 8.8 & 2 & 10.7 & $1.8\times10^{-10}$ & $2.2\times10^{33}$ & 14,15,4 \\
1H 1905+000 & $1.9\times10^{21}$ & $<50$ & 10 & - & - & $<1.1\times10^{-10}$ & $<4.8\times10^{31}$ & 16,15 \\
SAX J1808.4--3658 & $1.3\times10^{21}$ & $<34$ & 3.5 & 5 & 10.7 & $1.0\times10^{-11}$ & $<1.1\times10^{31}$ & 17,4,15 \\

\enddata 
\tablecomments{Estimates of quiescent thermal luminosities from 
neutron star transients, and mass transfer rates (inferred from RXTE ASM 
observations for systems with RXTE-era outbursts).  Quiescent thermal luminosities are computed for the unabsorbed NS component in the 0.01-10 keV range. Outbursts and years columns give the number of outbursts and the time baseline used to compute \.M, if this calculation was performed in this work (indicated by referring to reference 4).
References as follows: 1: \citet{Rutledge01b}, 
2: \citet{Campana03}, 
3: \citet{Tomsick04}, 
4: Mass transfer rate computed in this work,  
5: \citet{Rutledge01a},     
6: \citet{Rutledge99},
7: \citet{Cackett06b},
8: \citet{Tomsick05},
9: \citet{Wijnands05a},
10: \citet{Heinke06b},
11: \citet{Cackett05},
12: \citet{Cackett06a},
13: \citet{Jonker04}, 14: \citet{Jonker04b}, 
15: Quiescent bolometric luminosity computed in this work, 
16: \citet{Jonker06}, 
17: \citet{Galloway06}.
}
\end{deluxetable}





\end{document}